\documentclass[aps,prl,twocolumn,showpacs,groupedaddress]{revtex4}

\usepackage[dvips]{graphicx}
\usepackage{amssymb}
\usepackage{verbatim}
\usepackage{color}
\newcommand{\T}{{\cal T}}
\newcommand{\R}{{\cal R}}
\newcommand{\V}{{\cal V}}

\begin{document}
\title{Direct measurement of the coherence length
of edge states in the Integer Quantum Hall Regime}
\author{Preden Roulleau}
\author{F. Portier}
\author{P. Roche}
\email{patrice.roche@cea.fr}
\affiliation{Nanoelectronic group, Service de Physique de l'Etat Condens\'e,\\
CEA Saclay, F-91191 Gif-Sur-Yvette, France}
\author{A. Cavanna}
\author{G. Faini}
\author{U. Gennser}
\author{D. Mailly}
\affiliation{CNRS, Phynano team, Laboratoire de Photonique et Nanostructures,\\
Route de Nozay, F-91460 Marcoussis, France}
\date{\today}

\begin{abstract}
We have determined the finite temperature coherence length of edge
states in the Integer Quantum Hall Effect (IQHE) regime. This was
realized by measuring the visibility of electronic Mach-Zehnder
interferometers of different sizes, at filling factor 2. The
visibility shows an exponential decay with the temperature. The
characteristic temperature scale is found inversely proportional
to the length of the interferometer arm, allowing to define a
coherence length $\l_\varphi$. The variations of $\l_\varphi$ with
magnetic field are the same for all samples, with a maximum
located at the upper end of the quantum hall plateau. Our results
provide the first accurate determination of $\l_\varphi$ in the
quantum Hall regime.
\end{abstract}

\pacs{03.65.Yz, 73.43.Fj, 73.23.Ad} \maketitle The understanding
of the decoherence process is a major issue in solid state
physics, especially  in view of controlling entangled states for
quantum information purposes. The edge states of the quantum Hall
effect are known to present an extremely long coherence length
$l_\varphi$ at low temperature \cite{Martin90PRL64p1971},
providing a useful tool for quantum interference experiments
\cite{Ji03Nature422p415,Samuelsson04PRL92n026805,Litvin07PRB75n033315,Neder07Nature448p333,Roulleau07PRB76n161309}.
Surprisingly, very little is known on the exact value of this
length and the mechanisms which reduce the coherence of edge
states. This is in strong contrast with diffusive conductors,
where weak localisation gives a powerful way to probe $l_\varphi$.
It has been shown, in this case, that electron-electron
interactions are responsible for the finite coherence length at
low temperatures. In the IQHE regime, the presence of a high
magnetic field destroys any time reversal symmetry needed for weak
localisation corrections, making such an investigation difficult.
Furthermore, due to the uni-dimensionality of the edge states,
electron-electron interactions may strongly modify the single
particle picture and one can ask wether the notion of phase
coherence length is still relevant and how it depends on
temperature. In this letter, we show for the first time that one
can define a phase coherence length, and that it is inversely
proportional to the temperature.

\begin{figure}[h]
\centerline{\includegraphics[angle=-90,width=7.5cm,keepaspectratio,clip]{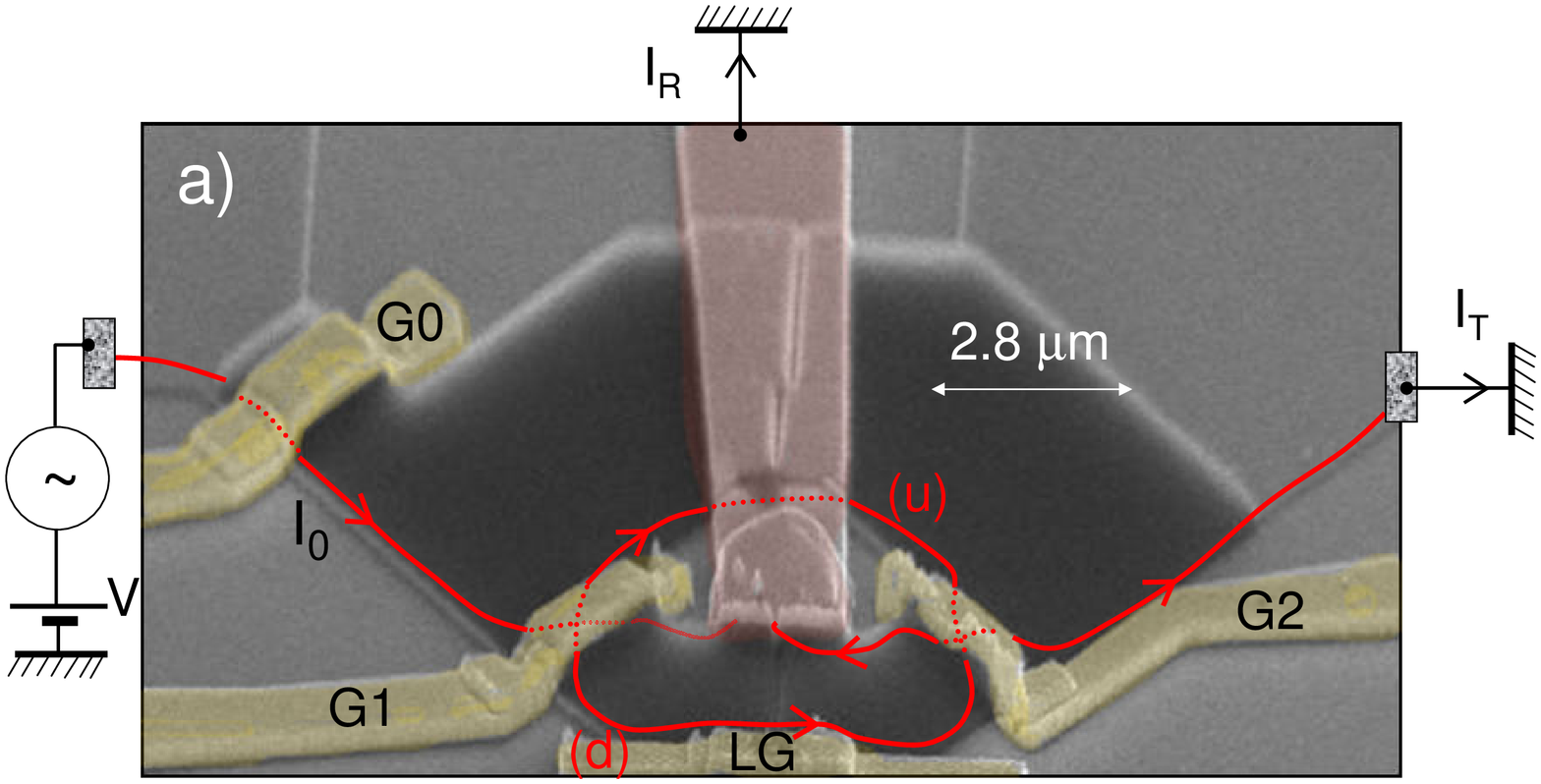}}
\centerline{\includegraphics[angle=-90,width=7.5cm,keepaspectratio,clip]{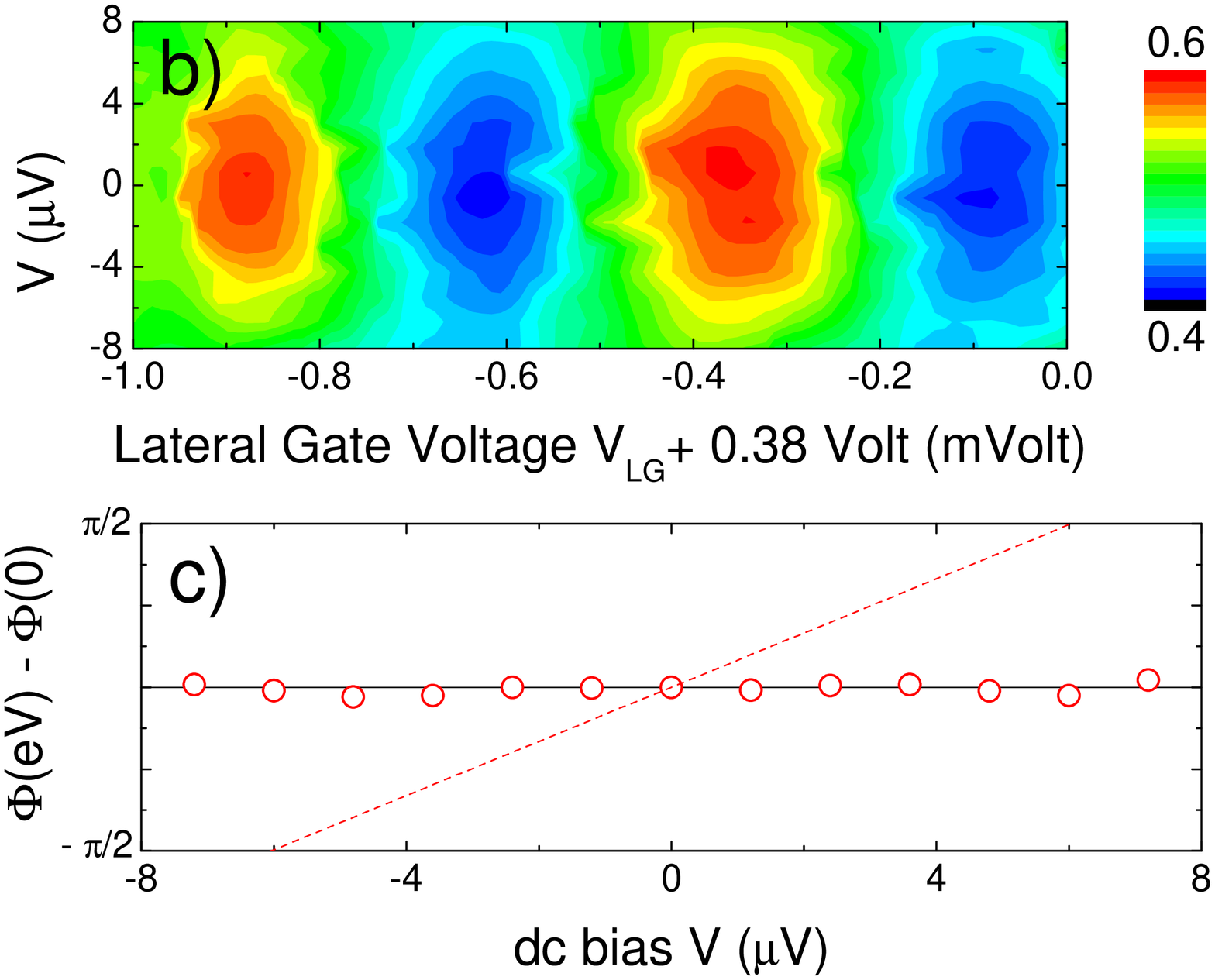}}
\caption{(color online) \textbf{a)} Tilted SEM view of the "small"
MZI. G0, G1 and G2 are QPCs whose split gates are connected with
gold bridges over an isolator responsible for the black color of
the SEM view. LG is a lateral gate which allows to vary the
surface defined by the two arms (u) and (d) of the MZI. The
interferences are measured in the IQHE at filling factor 2 on the
outer edge state represented by a line on the SEM picture. The
inner edge state is fully reflected by all the QPCs. The small
ohmic contact in between the two arms collects the back scattered
current $I_B$ to the ground through a long gold bridge.
\textbf{b)} 2D plot of $dI_T/dI_0$ as a function of the lateral
gate voltage $V_{LG}$ and the DC bias $V$, for the large sample at
$20$~mK. The visibility of interferences of the order of 20\%-40\%
decreases with $V$ while the phase of interferences remains almost
constant. \textbf{c)} Phase of the large sample deduced from Fig.
(1b). The dashed line is the energy dependence of the phase which
would be necessary to explain our observed visibility decrease
with thermal smearing (see text).}\label{Principe.fig}
\end{figure}

Though the energy redistribution length has been studied in the
past \cite{Alphenaar90PRL64p677,Machida97SSC103p441},these
scattering experiments do not measure the phase coherence, which
requires observation of electron interference effects.  So far,
experiments have only been able to put a lower bound on
$l_\varphi$ at low temperatures
\cite{vanwees89PRL62p2523,Bird94PRB50p14983,Ji03Nature422p415,Yang05PRB71p113312}.
The electronic Fabry-P\'erot interferences occurring in ballistic
quantum dots have been used since the early days of mesoscopic
physics \cite{vanwees89PRL62p2523}. These first studies showed an
exponential decay of the amplitude of the Aharonov-Bohm (AB)
oscillations with temperature\cite{Bird94PRB50p14983}. However,
this decay was attributed to thermal smearing due to the
contribution of thermally activated one particle energy levels of
the dot. Furthermore, the size of the interferometers was not
varied, nor was a Fourier analysis performed  of the AB
oscillations that could yield an estimation of $\l_\varphi$
\cite{Hansen01PRB64n045327}. Quantum dot systems also implicate
the possible interplay of Coulomb Blockade effects
\cite{Rosenow07PRL98n106801}. The Mach-Zender interferometers
(MZI)
\cite{Ji03Nature422p415,Litvin07PRB75n033315,Roulleau07PRB76n161309}
used in the present study do not suffer from the same limitations.
First, we will show that the observed oscillations result from the
interference of two paths of equal length, making thermal smearing
negligible. Second, charge quantization effect leading to  Coulomb
blockade are irrelevant here. Last, comparison between
interferometers of various sizes allows us the unambiguous
determination of $\l_\varphi$, as well as its dependence with
temperature and magnetic field.

The sample geometry, presented in Fig.~(\ref{Principe.fig}), is
the same as in \cite{Roulleau07PRB76n161309}. MZIs of different
sizes were patterned using e-beam lithography on a high mobility
two dimensional electron gas formed at the GaAs/Ga$_{1-x}$Al$_x$As
heterojunction (sheet density $n_S=2.0\times10^{11}$~cm$^{-2}$ and
mobility = $2.5\times10^{6}$~cm$^2$/Vs). The experiments were
performed in the IQHE regime at filling factor $\nu=n_Sh/eB=2$
(magnetic field $B\simeq$4.6 Tesla). Transport occurs through two
edge states. Quantum point contacts (QPC) G0, G1 and G2 define
electronic beam splitters with transmissions $\T_i$ (i=0-2). In
all the results presented here, the interferences were studied on
the outer edge state schematically drawn as red lines in
Fig.(\ref{Principe.fig}), the inner edge state being fully
reflected by all the QPCs. The first gate G0 is tuned to fully
transmit the outer ($\T_0$=1) edge state. The interferometer
itself consists of G1, G2 and the small central ohmic contact in
between the two arms. G1 splits the incident beam into two
trajectories (u) and (d), which are recombined with G2, leading to
interferences. Samples have been designed such that (u) and (d)
are of equal length. The sizes of the three interferometers used
in this study scale by up a factor $\sqrt{2}$: the length of their
arms are L = 5.6~$\mu$m, 8~$\mu$m and 11.3~$\mu$m for enclosed
areas of 8.5~$\mu$m$^2$ (referred to as 'small'), 17~$\mu$m$^2$
('medium') and 34~$\mu$m$^2$ ('large'), respectively. The samples
are cooled in a dilution refrigerator to temperatures ranging from
20~mK to $\sim$ 200~mK.

\begin{figure}
\centerline{\includegraphics[angle=-90,width=7.5cm,keepaspectratio,clip]{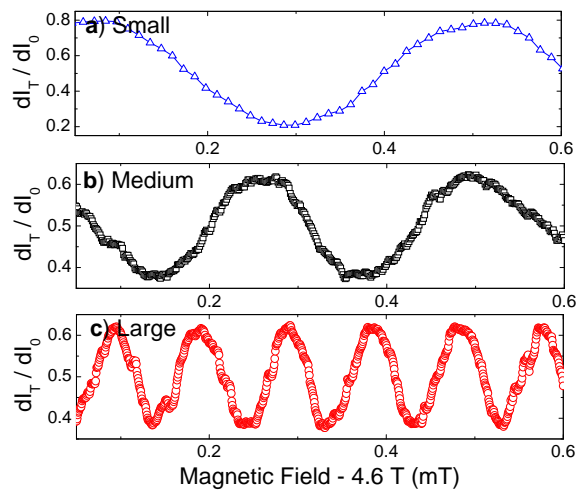}}
\caption{(color online) Interferences revealed upon varying the
magnetic flux through the surface defined by the two arms (u) and
(d) of the interferometers. From the oscillation period $\delta B$
we deduce the surface $S=h/(e\delta B)$ of the 3 different studied
MZI. \textbf{a)} The small MZI ($S=8.7 \pm 0.2\,\mu$m$^2$ ).
\textbf{b)} The medium  MZI ($S=15.5 \pm 0.4\,\mu$m$^2$ ).
\textbf{c)} The large MZI ($S=40.7 \pm 0.8\,\mu$m$^2$ ). All these
surfaces are in good agreement with the lithographic ones (see
text).}\label{oscB.fig}
\end{figure}

The labels are indicated in the upper part of
Fig.~(\ref{Principe.fig}). A current $I_0$ is injected into the
outer edge state through the interferometer. The current which is
not transmitted, $I_B=I_0-I_T$, is collected to the ground with
the small central ohmic contact. $I_0$ is made up of a minute AC
part, with the possibility to superimpose a DC bias $V$. The
differential transmission of the interferometer is defined as $\T=
G/G_0 =dI_T/dI_0$, where $G=dI_T/dV$ is the differential
conductance and $G_0=e^2/h$. It is measured with a standard
lock-in technique using a 619 Hz frequency and a 39 pA$_{rms}$
amplitude
 AC bias. The corresponding bias voltage excitation (1~$\mu$V$_{rms}$)
is always smaller than the energy scale involved. The oscillations
revealing the quantum interferences can be obtained using two
equivalent experimental procedures: either by superimposing a
minute current to the large current of the magnet, or by changing
the surface defined by the MZI using a lateral gate.
Fig.~(\ref{oscB.fig}) shows the AB oscillations of the
transmission for the three interferometers, showing a magnetic
period inversely proportional to the area of the interferometer,
while Fig.~(1b) shows oscillations obtained using the lateral gate
LG. After checking that both methods lead to the same
interferences amplitude, we have always used the lateral gate and
run the magnet in the permanent-current mode, strongly reducing
the measurement noise. The visibility of the AB oscillations is
defined as the ratio of the half amplitude of the oscillation of
the transmission divided by the mean value,
$\mathcal{V}=(\T_{\mathrm{max}}-\T_{\mathrm{min}})/(\T_{\mathrm{max}}+\T_{\mathrm{min}})$.

\begin{figure}
\centerline{\includegraphics[angle=-90,width=7.5cm,keepaspectratio,clip]{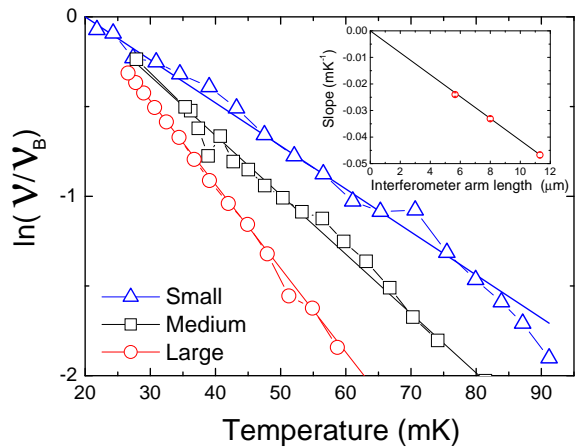}}
\caption{(color online) $\ln(\V/\V_B)$ as a function of the
temperature for the 3 different samples, $\V_{B}$ being the
visibility measured at $T_B$=20~mK. The measurement has been done
at the magnetic field  for which the visibility decay is the
smallest. \textbf{Inset} :  The slope $\ln(\V/\V_{B})/(T-T_B)$ is
proportional to the arm length.}\label{expdecay.fig}
\end{figure}

The maximum value of $\mathcal{V}$ is always obtained at the
lowest temperature. $\mathcal{V}$ can reach 65 \% for the small
interferometer, whereas it typically attains 20-40 \% for the
medium and the large interferometers (see Fig.~(1b)). For each
interferometer we have studied the temperature dependence of the
visibility. In fig.~(\ref{expdecay.fig}), we have plotted
$\ln(\mathcal{V}/\mathcal{V}_{B}$) versus temperature, where
$\mathcal{V}_{B}$ stands for the visibility at $T_B$=20~mK.
Clearly, the visibility decreases with temperature in all cases,
and the larger the interferometer, the stronger the temperature
dependence. More quantitatively, if a linear regression of
$\ln(\mathcal{V}/\mathcal{V}_{B})=(T-T_B)/T_0$ is done, one finds
that $T_0^{-1}$ is proportional to the length of the interfering
arms (inset of Fig.~(\ref{expdecay.fig})). In the following, we
show that this behavior does not result from a thermal smearing.

The transmission probability through the MZI at the energy
$\epsilon$ is $\T(\epsilon) =
\T_1\T_2+\R_1\R_2+z\sqrt{\T_1\R_2\R_1\T_2}\sin[\phi(\epsilon)]$,
where $z\in[0,1]$ is a parameter accounting for phase averaging
and/or decoherence, and $\T_i=|t_i|^2=1-\R_i$ are the beam
splitters transmissions
\cite{Marquardt04PRB70n125305,Chung05PRB72p125320}.
$\phi(\epsilon)$ is the AB flux across the surface $S(\epsilon)$
defined by the energy dependent edge state positions in the two
interfering arms, $\phi(\epsilon)= 2\pi S(\epsilon)\times eB/h$.
When there is a finite length difference $\Delta L=L_u-L_d$
between the two arms, the surface $S$ depends on the energy
$\epsilon$. Thus the phase varies with the energy,
$\phi(\epsilon+E_F)=\phi(E_F)+\epsilon/(k_BT_S)$ where $k_BT_S=
\hbar v_D/\Delta L$ \cite{Chung05PRB72p125320} and $v_D$ is the
drift velocity (10$^4$ to 10$^5$
ms$^{-1}$)\cite{Ashoori92PRB45r3894}. The differential conductance
$G$ at bias $V$ and at temperature $T$ probes the transmission
probability at energy $eV$ smeared over an energy range $k_BT$
\cite{Chung05PRB72p125320}: $G(V)=G_0\int_{-\infty}^{+\infty}
f(\epsilon)\T(\epsilon+eV)\mathrm{d}\epsilon \propto [1+\V_0
<\sin(\phi(eV)>_{k_BT})],$ where $f(\epsilon)$ is the Fermi
distribution at temperature $T$. The energy dependence of the
phase $\phi$ leads to a thermal smearing at finite temperature as
the phase is blurred. A complete calculation yields a visibility
decreasing like $\V=\V_0\pi T /(\pi T_S\sinh(\pi T/T_S))$
\cite{Chung05PRB72p125320}. At low temperature, $T_S$ can be
determined by measuring the phase of the interferences as a
function of the dc bias $V$: $\phi(eV)=\phi(0)+eV/(k_BT_S)$.

In order to fit the visibility decrease with thermal smearing,
this requires that $T_S \sim 66 , 59$ and 44~mK, for the small,
medium and large sample respectively. In Fig. (1b) we have plotted
a 2D graph of the differential transmission $\T(V)$ as a function
of the lateral gate voltage and the dc bias, for the large sample
at 20~mK. From this measurement we have deduced the phase
$\phi(eV)$ which is shown to remain almost constant over an energy
range of $\sim16$~$\mu$eV (Fig.~(1c)). As a comparison, the dashed
line of (Fig.~(1c)) is the phase dependence which would be
required ($T_S$=44~mK) to explain the decrease of the visibility
with thermal smearing. The conclusion is straightforward: our
sample does not suffer from thermal smearing. We have done the
same procedure for all the three samples which exhibits a phase
rigidity over at least $\sim16$~$\mu$eV, meaning that all our
samples have negligible thermal smearing in the explored
temperature range $k_BT < 16$~$\mu$V$\equiv$200~mK.

\begin{figure}
\centerline{\includegraphics[angle=-90,width=7.5cm,keepaspectratio,clip]{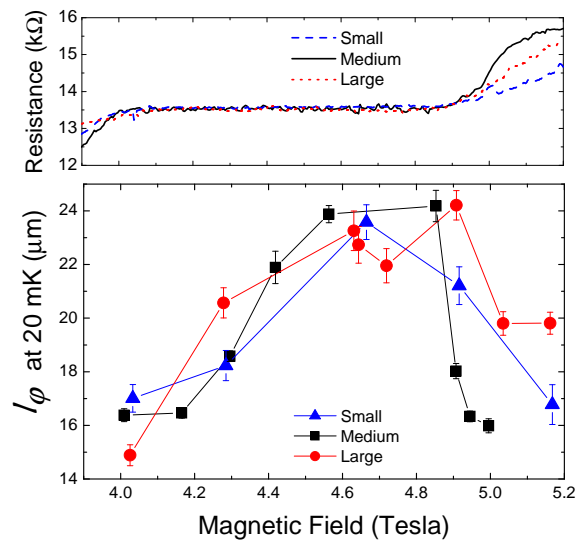}}
\caption{(color online) \textbf{Upper panel} : The dashed, solid
and dot lines are the two point Hall resistance at filling factor
2 measured for the small, the medium and the large sample
respectively. $l_\varphi$ has a general shape recovered by all the
three samples, with a maximum at the end of the Hall
plateau.\textbf{ Lower panel} : Coherence length at 20~mK deduced
from $L\times(T-T_B)/\ln(\V/\V_{B})$ for the three different
samples studied ($L$= 5.6 , 8 and 11.3 $\mu$m). The magnetic
fields (x-axis) of the small and large sample has been shifted by
+0.25 and -0.1 Tesla respectively, such that the plateau centers
coincide for the three samples.
}\label{renorm.fig}
\end{figure}

The exponential decrease of the visibility with temperature is
robust against various parameter variations, revealing a universal
behaviour. While the maximum visibility at the lowest temperature
is affected by varying the transmissions $\T_i$ of the MZI and by
applying a finite bias \cite{Roulleau07PRB76n161309}, the slope
$\ln(\V/\V_B)/(T-T_B)$ is found to be unaffected.

Indeed, and this is the central result of our paper, our
measurements can be interpreted by the introduction of a coherence
length $l_\varphi(T)$ such that
\begin{equation}
\V=\V_0 e^{-2L/l_\varphi} \;\;\mathrm{with}\;\; l_\varphi\propto
T^{-1}
\end{equation}
as shown in Fig.~(\ref{expdecay.fig}). $\V_0$ contains the
temperature independent part of the visibility. In the inset of
the Fig.~(3), we have plotted the slope $\ln(\V/\V_B)/(T-T_B)$ for
the three samples \footnote{These slopes are the minimum  values
obtained at different magnetic fields. From small to large MZI:
4.41 T, 4.91 T and 4.85 T.}. It is clear that the slope scales
with the length of the interferometer arm defining, \textit{de
facto}, a coherence length $l_\varphi(T)$ of  about 20~$\mu$m at
20~mK. The magnetic field variation of the deduced $l_\varphi$ is
independent from the MZI size (see Fig.~(\ref{renorm.fig})).  In
order to compare  the three samples at the same filling factor, we
have shifted the x-axis of Fig.~(\ref{renorm.fig})  by +0.25 T and
-0.1 T for the small and large MZI respectively. These values
center the Hall plateaus all together. The maximum of the
coherence length is reached at the upper end of the plateau where
the longitudinal resistance is usually minimum. However, our
sample configuration does not allow us to check if it is actually
the case.

Let us now compare our results with previously available data from
other groups. Although the variations of $\V(T)$ were measured
only for one interferometer size in the following experiments, it
is possible to fit the data with equation 1 and to deduce a
coherence length  value at 20~mK. In the Fabry-P\'erot type
interferometer (Fig.~(5b) of ref.\cite{Bird94PRB50p14983}), our
analysis, using a coherence length instead of thermal smearing,
leads to $l_\varphi\sim$20~$\mu$m at 20~mK. Although the
experiments of ref.\cite{Bird94PRB50p14983} were performed with
different filling factor, magnetic field, mobility, density and
geometry, surprisingly it gives the same result. The data from
ref.\cite{Ji03Nature422p415} yields also a similar $l_\varphi$,
although a direct comparison is difficult without an exact
knowledge of the MZI dimensions. Finally, the results of
ref.\cite{Litvin07PRB75n033315}, again interpreted by the authors
as resulting from thermal smearing, lead to
$l_\varphi\sim$~80~$\mu$m at 20~mK.

What kind of mechanism is responsible for a finite coherence
length varying with a $T^{-1}$ temperature dependence? Electron -
electron collisions are known to limit the coherence in non
unidimensional conductors (2D electron gas, diffusive metallic
conductors ...). For the MZI, a finite $l_\varphi$ coming from
either short range interaction ($l_\varphi\propto T^{-3}$), long
range interaction ($l_\varphi\propto T^{-1}\ln^2(1/T)$) or
curvature of the fermion dispersion ($l_\varphi\propto T^{-2}$)
\cite{Chalker07PRB76p085320} cannot explain our findings.
Alternatively, interactions with environment electrons,
capacitively coupled to the arms of the interferometer, have been
proposed to describe the decoherence of MZIs
\cite{Seelig01PRB64n245313}. More specifically, decoherence is due
to the thermal noise of the dissipative part of the finite
frequency coupling impedance between the environment and the
reservoirs. This theory leads to
\begin{equation}
\frac{l_\varphi}{L} =\frac{\tau_\varphi }{\tau}=\frac{3\hbar
}{2\pi k_B T}\frac{v_D}{L}\times \left( 1+\frac{\pi\hbar
Cv_D}{Le^2}\right)^2,
\end{equation}
up to a numerical factor decreasing from 3/2 to 1 when $\pi\hbar C
v_D /( L e^2)$ increases from 0 to $\infty$ (Eq. 37 of
ref.\cite{Seelig01PRB64n245313}). $C$ is a geometric capacitance,
which represents the coupling to the environment and $\tau$ is the
time of flight. For $v_D=5.10^4$~ms$^{-1}$ and
$C/L\sim\epsilon_r\epsilon_0$, one finds $\pi\hbar C v_D /( L e^2)
\ll 1$ and $l_\varphi\sim14$~$\mu$m at 20~mK. This result agrees
with our measurements, although in the absence of an independent
determination of $v_D$ and $C$, it is not possible to be more
quantitative.

We now turn to the non monotonic dependence of $l_\varphi$ with
the magnetic field $B$. If $\tau_\varphi$ is independent of $B$,
as suggested by equation (2), the apparent variation of
$l_\varphi$ results from a variation of $\tau$. As we have deduced
$l_\varphi$ assuming a constant trajectory length $l=L$, any
variation of $l$, due to disorder, would modify the deduced
$l_\varphi=\tau_\varphi L \times v_D/l$. Then, the maximum of
$l_\varphi$ shown in Fig.~(4) corresponds to the minimum of
$l/v_D$. In a naive picture, the drift velocity $v_D$ varies like
$1/B$ \cite{Beenakker90PRL64p216,Ashoori92PRB45r3894} barely
leading to a non monotonic variations of $l_\varphi$. In another
hand, a non monotonic variation of $l$ is all the more plausible.
The maximum of $l_\varphi$ occurs on the upper end of the Hall
plateau where one expects minimum backscattering, thus a minimum
$l$. Assuming this explanation correct, the overlap of the three
curves in Fig.~(4) (lower panel) indicates that the variations of
$l$ scale with the geometric length of the MZI. The study of the
influence of the sample disorder on the coherence length and its
dependence with magnetic field could bring new insights supporting
our assumption.

In conclusion, we have measured the visibility of Mach-Zehnder
interferometers of various sizes, operating in the IQHE regime at
filling factor 2, as a function of both the temperature and the
magnetic field. Our results provide a direct and reliable
measurement of the coherence length found to be inversely
proportional to the temperature and maximum at the upper end of
the Hall plateau. The order of magnitude is compatible with
theoretical predictions based on a dephasing arising from the
thermal noise of the environment.


\end{document}